\documentclass[prb,twocolumn,amsmath,superscriptaddress,amssymb,showpacs]{revtex4}

\usepackage{color} 
\usepackage{graphicx}
\usepackage{epstopdf}
\usepackage{bm}

\begin{document}

\DeclareGraphicsExtensions{.eps, .eps}
\bibliographystyle{prsty}

\title {Optical conductivity of V$_4$O$_7$ across its metal-insulator transition}

\author{I. Lo Vecchio}
\affiliation{Dipartimento di Fisica, Universit\`{a} di Roma "La Sapienza", Piazzale A. Moro 2, I-00185 Roma, Italy}

\author{M. Autore}
\affiliation{INFN and Dipartimento di Fisica, Universit\`{a} di Roma "La Sapienza", Piazzale A. Moro 2, I-00185 Roma, Italy}

\author{F. D'Apuzzo}
\affiliation{Istituto Italiano di Tecnologia and Dipartimento di Fisica, Universit\`{a} di Roma "La Sapienza", Piazzale A. Moro 2, I-00185 Roma, Italy}

\author{F. Giorgianni}
\affiliation{INFN and Dipartimento di Fisica, Universit\`{a} di Roma "La Sapienza", Piazzale A. Moro 2, I-00185 Roma, Italy}

\author{A. Perucchi}
\affiliation{Elettra-Sincrotrone Trieste S.C.p.A. and INSTM UdR Trieste, S.S. 14 Km 163,5 in AREA Science Park IT-34149 Basovizza, Trieste, Italy}

\author{U. Schade}
\affiliation{Helmholtz-Zentrum Berlin f\"ur Materialien und Energie GmbH Elektronenspeicherring BESSY II, Albert-Einstein-Strasse 15, D-12489 Berlin, Germany}

\author{V. N. Andreev}
\affiliation{Ioffe Physical-Technical Institute, Russian Academy of Sciences, Politekhnicheskaya ul. 26, St. Petersburg, 194021 Russia}

\author{V. A. Klimov}
\affiliation{Ioffe Physical-Technical Institute, Russian Academy of Sciences, Politekhnicheskaya ul. 26, St. Petersburg, 194021 Russia}

\author{S. Lupi}
\affiliation{CNR-IOM and Dipartimento di Fisica, Universit\`a di Roma "La Sapienza", Piazzale A. Moro 2, I-00185, Roma, Italy}
\date{\today}

\begin{abstract}
The optical properties of a V$_4$O$_7$ single crystal have been investigated from the high temperature metallic phase down to the low temperature antiferromagnetic insulating one. The temperature dependent behavior of the optical conductivity across the metal-insulator transition (MIT) can be explained in a polaronic scenario. Charge carriers form strongly localized polarons in the insulating phase as suggested by a far-infrared charge gap abruptly opening at T$_{MIT}$ $\approx$ 237 K. In the metallic phase instead the presence of a Drude term is indicative of fairly delocalized charges with a moderately renormalized mass \textit{m}$^*$$\approx$ 5\textit{m$_e$}. The electronic spectral weight is almost recovered on an energy scale of 1 eV, which is much narrower compared to VO$_{2}$ and V$_{2}$O$_{3}$ cases. Those findings suggest that electron-lattice interaction rather than electronic correlation is the driving force for V$_4$O$_7$ metal-insulator transition.
\end{abstract}
\maketitle

\section*{I. INTRODUCTION}
Vanadium oxides are the archetypal metal-insulator transition (MIT) systems and their properties have been triggering a huge attention for decades both theoretically and experimentally. Besides the popular V$_2$O$_3$ and VO$_2$, where the vanadium valency is three and four respectively, several mixed valence compounds have been synthesized giving rise to the vanadium-oxide Magn\'eli phase \cite{Eyert-04}. This series of compounds is defined by the formula V$_n$O$_{2n-1}$=V$_2$O$_3$+($n$-2)VO$_2$, where 3$\leq$$n$$\leq$9. 
Except for V$_7$O$_{13}$, all of them undergo a metal-insulator transition accompanied by structural transformations\cite{Baldassarre-07,Cava-13, Eyert-03, Eyert2-03}. Moreover a long-range antiferromagnetic ordering occurs at N\'eel temperatures T$_{N}$ much lower than T$_{MIT}$, thus suggesting that the insulating state is not related to the magnetic order.

In V$_4$O$_7$ the MIT occurs at the temperature T$_{MIT}$ $\approx$ 237 K with a resisitivity jump of one order of magnitude \cite{Andreev-09} and a small thermal hysteresis (less than 1 K) \cite{Okinaka-70}.  As a matter of fact the entire temperature range from room temperature to $\approx$ 150 K is the range of the continuous transition where resistivity changes over three orders of magnitude. The antiferromagnetic order instead appears at T$_{N}$$\approx$ 34 K. 
The crystal structure of V$_4$O$_7$ consists of rutile-like blocks (as in VO$_{2}$) extending in two dimensions and four VO$_6$ octahedra wide in the third dimension \cite{Marezio-72, Marezio-78}. The shear planes among rutile blocks recall the V$_{2}$O$_{3}$ local structure \cite{Dernier}. The resulting lattice is triclinic, belonging to a P$\bar1$ space group, and showing four and seven independent V and O sites respectively.
The V sites are split into two groups: V3 and V4 at the shear plane, V1 and V2 located at the center of the rutile blocks. While the environment of V1 and V2 is VO$_{2}$-like, the local structure is V$_{2}$O$_{3}$-like around V3 and V4 octahedra which are connected by face sharing. These cations form two independent chains, V3-V1-V1-V3 (3113) and V4-V2-V2-V4 (4224), running parallel to the pseudorutile $c$-axis. At variance with VO$_{2}$ and V$_{2}$O$_{3}$, this low-symmetry crystal structure does not change at the MIT, except for a small contraction of the cell volume reducing  from 441.33 \AA$^{3}$ in the metallic phase to 440.12 \AA$^{3}$ in the insulating one (at 120 K) \cite{Marezio-78}.
The electron count in V$_4$O$_7$ provides an average valence of V$^{3.5+}$ for the metallic cations. In presence of charge ordering the average valence of 3.5 should correspond to a 2V$^{3+}$ and 2V$^{4+}$ ion distribution $per$ formula unit. As a matter of fact, X-ray diffraction data \cite{Marezio-78} show that the metallic state is characterized by an almost complete metal cation disorder. On the other hand, data in the insulating state analyzed in terms of a bond-length summation method \cite{Marezio-78}, suggest a high degree of charge ordering, nearly corresponding (at 120 K) to a 83$\%$ of V$^{3+}$ and V$^{4+}$ ordered cations along the 3113 and 4224 chains respectively. A Local Density Approximation (LDA)+U electronic band calculation (where U is the local electron Hubbard repulsion of $d$-electrons in vanadium ions) is in agreement with this picture \cite{Botana-11} suggesting the importance of charge ordering in establishing the MIT and the insulating phase of V$_4$O$_7$.

Despite the interesting behavior of V$_4$O$_7$ unveiling aspects related to the physics of both VO$_{2}$ and V$_{2}$O$_{3}$, its properties have only been investigated with transport measurements\cite{Andreev-09, Kacki} to the best of our knowledge. These data suggest a strong role played by electron-lattice interaction in driving the MIT. Moreover, the temperature dependence of the dc conductivity in the insulating phase has been explained in terms of a small polaron hopping among V$^{3+}$ and V$^{4+}$ ordered lattice sites \cite{Andreev-09}. 
However a complete understanding of V$_4$O$_7$ electrodynamics in a broad frequency range is still lacking in literature. In this paper we fill this gap by presenting an optical study from the far-infrared to the visible range of a high quality V$_4$O$_7$ single crystal. We experimentally demonstrate that the optical conductivity in the metallic state of V$_4$O$_7$ is characterized by mobile charge carriers having a moderately renormalized mass \textit{m}$^*$$\approx$ 5\textit{m$_e$} due to a polaronic effect. A charge gap abruptly opens at the MIT in concomitance with a strong spectral weight transfer from the Drude term to a mid-infrared absorption. This suggests that at the MIT mobile polarons transform in strongly localized ones in agreement with diffraction data.

\section*{II. METHODS}
A stoichiometric single crystal of V$_4$O$_7$ was synthesized from a mixture of V$_2$O$_5$ and V$_2$O$_3$ (1:3) through gas transport in TeCl$_4$ vapors at a temperature of $\approx$ 1300 K. An X-ray diffraction analysis confirmed the absence of other vanadium oxide phases in the crystals. The sample showed a well reflecting mirror-like surface, thus no polishing process was used to perform reflectivity measurements. The near-normal incidence reflectance R($\omega$) data from the far-infrared to the visible range were collected using a Bruker Michelson interferometer and a helium flux optical cryostat working at a vacuum better than 10$^{-6}$ mbar.  BESSY-II storage ring radiation was exploited for the low-frequency part due to the small size of the crystal. A gold or silver (depending on the spectral range) coating was evaporated in situ over the sample surface and used as a reference.  
Measurements were perfomed between 10 and 300 K, thus allowing to probe all the phases that the sample undergoes, from the antiferromagnetic state to the metallic one.
Low-frequency reflectance data were extrapolated with standard methods (Hagen-Rubens or constant lines). Since no temperature dependence is found above 15000 cm$^{-1}$, the same standard 1/$\omega^4$ high-frequency extrapolation was used both for the insulating and the metallic phase. The complex optical conductivity $\sigma_1(\omega)$+i$\sigma_2(\omega)$ of the sample was thus obtained through Kramers-Kronig transformations. Due to the triclinic lattice structure, the reflectance was measured with unpolarized radiation and the optical conductivity cannot be related to a specific axis of the crystal.

\section*{III. RESULTS AND DISCUSSION}
The reflectance R($\omega$) of V$_4$O$_7$ is shown on a logarithmic scale for selected temperatures in the upper panel of Fig. 1. In the insulating phase from 10 to 200 K it is nearly constant at $\approx$ 0.5 for $\omega$ approaching zero frequency and then shows several phonon lines in the far-infrared range 50-800 cm$^{-1}$. At 230 K R($\omega$) increases to 0.7 and the intensity of the vibrational peaks is fairly reduced. Crossing T$_{MIT}$ $\approx$ 237 K the reflectance becomes metallic like approaching 1 at low frequencies. All curves (both in the insulating and in the metallic phases) merge around 8000 cm$^{-1}$.

\begin{figure}[t]
\leavevmode
\centering
\begin{center}
\includegraphics[width=8.5cm]{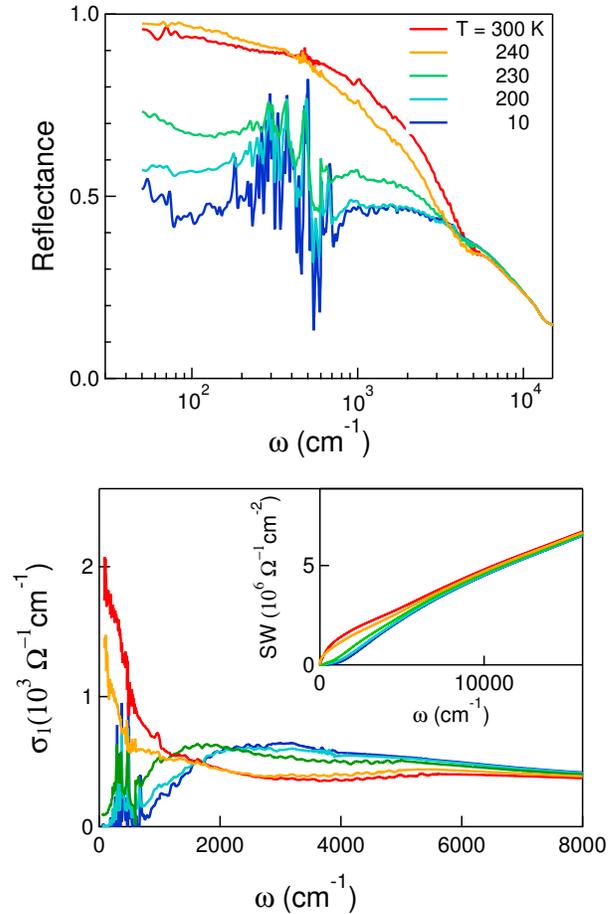}  
\end{center}
\caption{Upper panel: Near-normal incidence reflectance data of V$_4$O$_7$ plotted on a log scale for selected temperatures. Lower panel: real part of the optical conductivity as obtained by Kramers-Kronig transformations. The transition temperature is T$_{MIT}$ $\approx$237 K. The spectral weight distribution as obtained by integrating the real part of the optical conductivity (Eq. 1 in the text) is shown in the inset.}
\label{Fig1}
\end{figure}

The real part of the optical conductivity $\sigma_1(\omega)$ is shown in the bottom panel of Fig.1 on a linear scale up to 8000 cm$^{-1}$. In the insulating phase it shows a gap region followed by a mid-infrared (MIR) band centered at about 2500 cm$^{-1}$, varying little up to 230 K, where it starts to shift towards lower frequencies. In the metallic phase this band is strongly depleted in favor of a Drude metallic contribution. The main temperature dependent changes of $\sigma_1(\omega)$ occur below 8000 cm$^{-1}$ (1 eV), as it is pointed out by the behavior of the spectral weight (SW)
\begin{equation}
SW(\Omega,T)=\int_{0}^{\Omega}\sigma_1(\omega,T)\mathrm{d}\omega
\label{SW}
\end{equation}
(shown in the inset of Fig.1), proportional to the number of carriers taking part to the optical absorption up to a cutoff frequency $\Omega$. This marks a major difference with respect to VO$_{2}$ and V$_{2}$O$_{3}$ where the MIT, which is mainly due to electronic correlations,  induces a SW redistribution covering a broader spectral range up to the near-ultraviolet \cite{Basov_VO2, Baldassarre-08, Lupi-NC}. This behavior of the SW may imply that
charge localization and ordering in V$_4$O$_7$ is mainly induced by the
electron-lattice interaction.

In order to take a deeper look at the MIT, we will separately discuss the insulating and the metallic phases of V$_4$O$_7$ in the next two subsections.

\subsection*{III A. Insulating phase}
The optical conductivity in the insulating phase is characterized by a far-infrared gap which starts to open when approaching T$_{MIT}$ $\approx$ 237 K (see the bottom panel of Fig.1). The opening of this gap accompanies the appearance of a broad MIR absorption peaked around 1500 cm$^{-1}$ near T$_{MIT}$ and hardening to 2500 cm$^{-1}$ at 10 K. We estimated the width of the insulating gap by fitting the edge of the mid-infrared absorption with the following equation:

\begin{equation}
\sigma_{1}(\omega)=A[\omega-E_g(T)]^\alpha \hspace{0.5cm}\mbox{for}\hspace{0.2 cm} \omega\geq E_g
\label{parabolic}
\end{equation}

\begin{figure}[t]
\leavevmode
\centering
\begin{center}
\includegraphics[width=8.5cm]{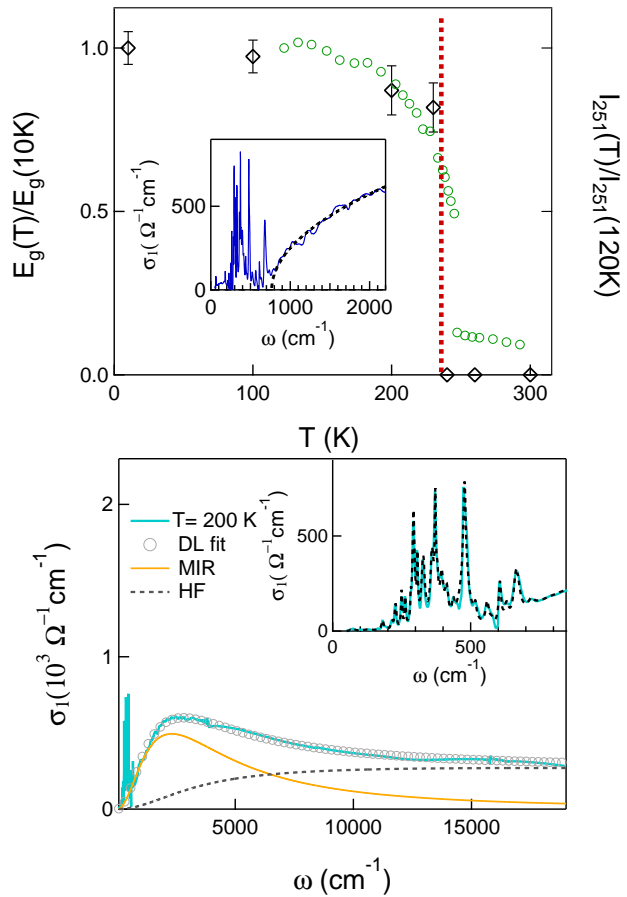}  
\end{center}
\caption{Upper panel: Temperature evolution of the energy gap (empty diamonds) showing a jump at T$_{MIT}$ $\approx$ 237 K, thus confirming the first order nature of the MIT. A standard fit of the gap edge (see text) was used to estimate the E$_g$(T) values as shown by a dashed line in the upper inset for T=100 K. In the right axis the temperature behavior of the (251) Bragg peak intensity (empty circles) is reported \cite{Marezio-72}. Lower panel: Drude-Lorentz fit of the optical conductivity (open circles) in the insulating phase at 200 K. The MIR band is represented as a thin solid curve, while the high-frequency temperature independent tail as a dashed curve. In the lower inset the phonon spectrum has been fitted through 30 Lorentzian peaks in agreement with the expected number of phonon modes admitted by the P$\bar1$ crystal symmetry. Their shapes and number remain invaried crossing the N\'eel temperature.} 
\label{Fig2}
\end{figure}

Here we assume a behavior of $\sigma_1(\omega)$ similar to that of a semiconductor in the presence of direct band to band transitions \cite{Lo Vecchio-13}. The curves thus obtained, reported as a dashed line in the inset of Fig.2 (upper panel) for T=100 K, nicely fit the rising edge of the optical conductivity at all temperatures with $\alpha\approx$1/2. The value of the gap E$_g$(T) corresponds to the intersection between Eq. 2 and the frequency axis. The ratio E$_g$(T)/E$_g$(10 K) is plotted by empty diamond symbols in the upper panel of Fig.2. At 10 K the gap E$_g$(10 K) is approximately 800 cm$^{-1}$, then it smoothly decreases when raising the temperature up to 230 K. When crossing the transition temperature (vertical dotted line), the gap abruptly closes. This is a clear signature of the first-order nature of the transition. 

In order to associate this gap to the charge ordering process,  we report the behavior of the (251) X-ray diffraction peak intensity I$_{251}$(T)/I$_{251}$(120 K) $vs.$ T in the same figure with open circles. This peak, which shows a discontinuous variation at T$_{MIT}$, has been proved to be particularly sensitive to the establishing of charge ordering in V$_4$O$_7$ \cite{Marezio-72}. The temperature dependent behavior of E$_g$(T) looks very similar to that of I$_{251}$(T) clearly showing a  correlation between the charge order (as measured by diffraction) and the gap opening. This strongly suggests that the MIT as observed in the low-energy electrodynamics of V$_4$O$_7$ is determined by the charge ordering process. 

The optical conductivity can be also analysed in terms of a Drude-Lorentz (DL) model:

\begin{equation}
\tilde{\sigma}(\omega)=\frac{\omega_p^2\tau}{4\pi(1-i\omega\tau)}+\frac{\omega}{4\pi i}\sum_j\frac{S_j^2}{\omega_j^2-\omega^2-i\omega\gamma_j}
\label{Eq2}
\end{equation}

\noindent where $\omega_p$ is the plasma frequency, $\tau$ the scattering time, $S_j$ and $\gamma_j$ are the strength and the spectral width of the Lorentz oscillators peaked at finite frequencies $\omega_j$. 
In the insulating phase, plotted in the bottom panel of Fig.2 for T= 200 K, the Drude contribution is absent, and we used one Lorentzian oscillator to reproduce the MIR absorption at about 2500 cm$^{-1}$ and one overdamped oscillator centered around 15000 cm$^{-1}$ to describe the high-frequency temperature independent background. 

As discussed in the introduction, the insulating state corresponds to a V$^{3+}$ and V$^{4+}$ charge ordering in the 1221 and 4334 vanadium chains running  along the pseudorutile $c$-axis. The oxygen octhaedra surrounding the vanadium 3+ and 4+ ions show very different V-O distances. V$^{4+}$ cations, distributed in the 1331 chains, have an average V(1)-O (V(3)-O) distance of 1.946 \AA (1.959 \AA) at 120 K. The corresponding distances for the V$^{3+}$ cations in the 2442 chains are instead 1.989 \AA (2.009 \AA) for V(2)-O (V(4)-O). Therefore, the localized holes in the V(1)-O and V(2)-O octahedra should be more properly described in terms of small polarons. In this scenario, is then natural to associate the MIR absorption to the photo-induced hopping of those polarons among the V$^{3+}$ and V$^{4+}$ lattice sites. In other words, the energy cost for a 3+$\rightarrow$ 4+ polaron charge fluctuation in the charge ordered state is of the order of 0.3 eV. 
It is also worth noticing that the MIR band at 100 K (not shown) has the same characteristic energy of that at 10 K. This implies that the magnetic order at T$_{N}$ does not further localize the charge carriers, thus confirming its irrelevant role in the low energy electrodynamics of V$_4$O$_7$.

The MIR band observed in the insulating phase of V$_4$O$_7$ looks very similar to that measured in other charge ordered polaron materials like V$_3$O$_5$ \cite{Baldassarre-07} and Fe$_3$O$_4$ \cite{Gasparov}. In particular, the MIR bands in the insulating charge ordered states of V$_4$O$_7$ and V$_3$O$_5$ are centered around 2500 and 3200 cm$^{-1}$ respectively. This similarity suggests that the ordering of charge carriers in V$_4$O$_7$ is mainly due to electron-lattice interaction possibly supported by a further localization energy gain related to electronic correlation. 

The far-infrared part of the optical conductivity is characterized by strong phonon absorptions. The primitive cell of V$_4$O$_7$ (P$\bar1$ crystal symmetry) contains two formula units, corresponding to 22 atoms located at the Wyckoff positions 2i \cite{Marezio-78, Eyert-04}. This symmetry admits three acoustic modes, 33 Raman A$_g$ phonons and 30 A$_u$ infrared modes. $\sigma_1(\omega)$ (see the inset in the bottom panel of Fig.2) can be reasonably fitted through 30 Lorentz oscillators both above and below the magnetic transition at T$_{N}$=34 K. This further suggests, in agreement with the previous findings, that the magnetic ordering does not play an important role in the low energy electrodynamics of V$_4$O$_7$.

\subsection*{III B. Metallic phase}
The metallic optical conductivity is reported in Fig.3 from far-infrared to visible frequencies. $\sigma_1(\omega)$ at 300 K consists of a nearly flat background extending up the visible and an absorption in the infrared, particularly evident below 2000 cm$^{-1}$. The intensity of this term decreases approaching T$_{MIT}$ thus transferring spectral weight towards higher frequencies. A DL fit of the optical conductivity provides a Drude contribution (characterized by a $\omega_{p}$=7400 cm$^{-1}$ and a scattering rate $\Gamma$=450 cm$^{-1}$), plus two Lorentz oscillators, one in the mid-infrared at about 1500 cm$^{-1}$ and another centered aroud 15000 cm$^{-1}$. This last oscillator reproduces the high-frequency temperature independent background also present in the insulating phase.

\begin{figure}[t]
\leavevmode
\centering
\begin{center}
\includegraphics[width=8.5cm]{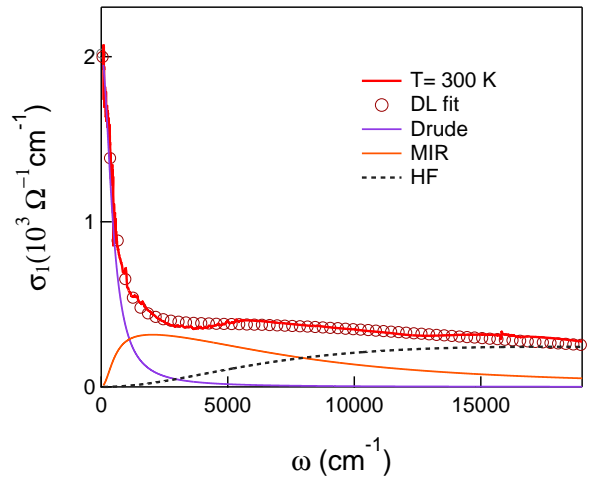}  
\end{center}
\caption{Drude-Lorentz fit of the optical conductivity in the metallic state (empty circles) at 300 K. The Drude and the MIR band are plotted as thin solid curves. The high-frequency temperature independent tail as a thin dashed line.}
\label{Fig3}
\end{figure}

The Drude term indicates a fairly coherent low energy electrodynamics in V$_4$O$_7$. This is at variance with V$_3$O$_5$ where most of the spectral weight in the metallic phase is located in the MIR range, indicating instead an incoherent charge transport. However, some degree of short range charge disproportion V$^{3+}$-V$^{4+}$ and octahedra distortion persists at room temperature in V$_4$O$_7$ as observed  by diffraction data \cite{Marezio-72}. Then, the low energy transport in the metallic phase could be related to mobile polarons rather than free charge carriers. Thus, their coherent propagation corresponds to the Drude behavior at low frequency, meanwhile their fast exchange among V$^{3+}$-V$^{4+}$ sites determines the MIR absorption at about 1500 cm$^{-1}$.
Indeed for a short-range charge ordering one can expect a lower energy cost for a charge fluctuation and this explains the softening of the MIR frequency across the MIT \cite{Ciuchi-99}. 

The MIR and Drude intensities at 300 K can be used to estimate the polaronic mass \textit{m}$^{*}$. Indeed, the ratio \textit{m}$^{*}$/\textit{m}$_{e}$=($\omega^{2}_{p}+S^{2}_{MIR})$/$\omega^{2}_{p}$$\approx$ 5, where \textit{m}$_{e}$ is the bare electron mass \cite{Van, Devreese, Perucchi-Nano}, suggests an intermediate electron-phonon coupling as observed, for instance, in lightly doped cuprates \cite{Lupi-PRB98}. This is at variance with the metallic phase of V$_3$O$_5$, where \textit{m}$^{*}$/\textit{m}$_{e}\approx$ 15 indicating less mobile polarons and an incoherent charge-transport.

\section*{IV. CONCLUSIONS}

In this paper we investigated the optical properties of a V$_4$O$_7$ single crystal across its metal to insulator transition at T$_{MIT}$ $\approx$ 237 K. 
The metallic optical conductivity is characterized by a mid-infrared absorption superimposed to a robust Drude term which indicates a fairly coherent charge transport. However, in agreement with X-ray diffraction measurements, a certain degree of local charge order associated with distorted V-O octahedra persists at room temperature. Thus, the low energy transport in the metallic phase could be related to mobile polarons rather than free charge carriers. The mid-infrared absorption then corresponds to the incoherent part of polaron conductivity and is probably related to fast V$^{3+}$/V$^{4+}$ charge fluctuations. 

The insulating state is instead characterized by a charge gap in the far-infrared which discontinuously  opens for T$<$T$_{MIT}$ in concomitance with the establishing of the charge ordering state as observed by X-ray diffraction. This gap accompanies a mid-infrared absorption at higher energy than the metallic state, representing the photoinduced hopping of strongly localized polarons among V$^{3+}$ to V$^{4+}$ long-range ordered lattice sites. 
All these experimental observations finally suggest that in V$_4$O$_7$ electron-lattice interaction is the mainly driving force for charge ordering, possibly supported by a further localization energy gain related to electronic correlation.

\section*{ACKNOWLEDGEMENTS}
We gratefully acknowledge discussions with Francesco Capitani. This study was supported by the Branch of Physical Sciences of the Russian Academy of Sciences within a research program. The research leading to these results has received funding from the European Community's Seventh Framework Programme (FP7/2007-2013) under grant agreement n. 312284.

\end{document}